\documentclass[useAMS,usenatbib]{mn2e}
\usepackage{graphics,graphicx,array,amssymb}

\title[Pulsation at the tip of the first giant branch?]{Pulsation at the tip of the first giant branch?}
\author[Y. Ita et al.]{Yoshifusa Ita$^{1}$\thanks{E-mail: yita@ioa.s.u-tokyo.ac.jp}, Toshihiko Tanab\'{e}$^{1}$, Noriyuki Matsunaga$^{1}$, Yasushi Nakajima$^{2}$,
\newauthor Chie Nagashima$^{2}$, Takahiro Nagayama$^{2}$, Daisuke Kato$^{2}$, Mikio Kurita$^{2}$,
\newauthor Tetsuya Nagata$^{2}$, Shuji Sato$^{2}$, Motohide Tamura$^{3}$, Hidehiko Nakaya$^{4}$
\newauthor and Yoshikazu Nakada$^{1,5}$
\\
$^1$Institute of Astronomy, School of Science, The University of Tokyo, Mitaka, Tokyo 181-0015, Japan\\
$^2$Department of Astrophysics, Nagoya University, Chikusa-ku, Nagoya 464-8602, Japan\\
$^3$National Astronomical Observatory of Japan, Mitaka, Tokyo 181-8588, Japan\\
$^4$Subaru Telescope, National Astronomical Observatory of Japan, 650 North A'ohoku Place, Hilo, HI 96720, U.S.A.\\
$^5$Kiso Observatory, School of Science, The University of Tokyo, Mitake, Kiso, Nagano 397-0101, Japan
}
\begin{document}

\date{Received 18 September 2002 / Accepted 10 October 2002}

\pagerange{\pageref{firstpage}--\pageref{lastpage}} \pubyear{2002}

\maketitle

\label{firstpage}

\begin{abstract}The first results of our ongoing near-infrared (NIR) survey of the variable red giants in the Large Magellanic Cloud, using the Infrared Survey Facility (IRSF) and the SIRIUS infrared camera, are presented. Many very red stars were detected and found that most of them are variables. In the observed colour-magnitude diagram ($J-K, K$) and the stellar $K$ magnitude distribution, the tip of the first giant branch (TRGB), where helium burning in the core starts, is clearly seen. Apart from the genuine AGB variables, we found many variable stars at luminosities around the TRGB. From this result, we infer that a substantial fraction of them are RGB variables.
\end{abstract}

\begin{keywords}
galaxies: Magellanic Clouds -- stellar content -- stars: AGB and post-AGB -- variables -- infrared: stars -- surveys
\end{keywords}

\section{Introduction}
After leaving the main sequence, stars of low to intermediate mass (0.8$M_\odot\lesssim M \lesssim8M_\odot$) become red giants and climb the first giant branch (RGB). At the tip of the RGB (TRGB), they experience the so-called helium flash, followed by a sudden decrease in luminosity. Then, the stars become red giants second time and ascend the asymptotic giant branch (AGB), increasing the scale of mass-loss and pulsation. On the upper AGB they become optically obscured because of the circumstellar dust shell formed by heavy mass-loss, making them accessible only in the infrared (\citealt{iben}).

\begin{figure}
\centering
\includegraphics[angle=0,scale=0.95]{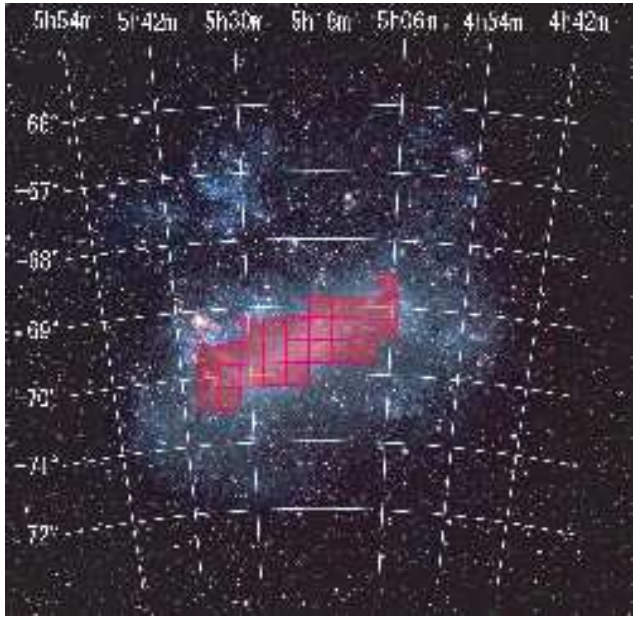}
\caption{A schematic illustration showing the survey region. The background image is a visible one and the coordinates are based on ICRS 2000 system. Each mesh has a side of 20 arcmin and consists of 9 sub-meshes, which correspond to the field of view of the SIRIUS.}
\label{map}
\end{figure}

Recently, large and homogeneous NIR photometry data for the Large Magellanic Cloud (LMC) has become available owing to large-scale NIR surveys such as 2MASS (\citealt{skrutskie}) and DENIS (\citealt{epchtein}). However, as they are single-epoch surveys, they do not provide variability information. Variability information is necessary to get accurate mean magnitudes that are also indispensable for a proper understanding of the red giant evolution.

Many optically-visible variables were found in the LMC, as the by-products of the gravitational lensing experiments like MACHO (\citealt{wood2000}), MOA (\citealt{noda}) and OGLE (\citealt{udalsky}). However, most, if not all, of high mass-losing variables might have been missed by these projects, because they are not readily detectable in the optical wave bands due to their surrounding circumstellar dust shells. Typical examples of such dust-enshrouded variables are OH/IR stars and dusty carbon stars (e.g., \citealt{habing1996}). As these dust-enshrouded variables are thought to be at the latest stage of AGB evolution (e.g., \citealt{habing1996}), the understanding of final evolution of low to intermediate mass star is greatly helped if their accurate quantities such as their number density and other physical parameters (mean magnitudes and periods etc.) would be obtained through deeper and larger survey.

Many low-amplitude variables were newly discovered by the aforementioned optical surveys. The vast majority are at the luminosities below the TRGB. \citet{alves} suggested that they are AGB stars in the early evolutionary phase prior to entering the thermal-pulsing phase of the AGB. However, it is not clear yet if they really are or not.

In this letter, we present some first results of our NIR monitoring survey which was started in December 2000, using the InfraRed Survey Facility (IRSF) at Sutherland, South African Astronomical Observatory. The limiting magnitudes (10$\sigma$) of this survey are 17.4, 17.0 and 15.8 at $J$, $H$ and $K$, respectively. At these limits, we can cover all of the AGB stars except extremely red ones (i.e., $J-K \gtrsim 5.0$, such as those found in NGC419 and NGC1978 by \citealt{tanabe1998}) and partway down the RGB populations in the LMC. We have been monitoring the total area of 3 square degrees along the LMC bar (see Figure.~\ref{map}), sufficiently large to do statistical analysis and to make complete catalog of variable red giants.

\section[]{Observations and reductions}
The IRSF consists of a specially constructed 1.4m alt-azimuth telescope to which is attached a large-format 3-channel infrared camera, SIRIUS (Simultaneous three-colour InfraRed Imager for Unbiased Surveys). The SIRIUS can observe the sky in the three wave bands $J$(1.25$\mu$m), $H$(1.63$\mu$m) and $K_s$(2.14$\mu$m) simultaneously and has a field of view of 7.8 arcmin square with a scale of 0.45 arcsec/pixel. Details of the instrument are found in \citet{nagashima} and \citet{nagayama}.

\subsection{Photometry}
Standard image reduction (i.e., dark subtraction, flat-field correction, sky subtraction) was made by using the SIRIUS pipeline (Nakajima, private communication). All images were reduced in the same way and a single image comprises 10 dithered 5 sec exposures. Photometry was performed on the dithered images with DoPHOT in the fixed position mode (\citealt*{schechter}).

At first, the instrumental magnitudes were transformed to the LCO system ones based on observations of a few dozen stars from Persson et al. (\citealt{persson}). Then, the LCO system magnitudes were further transformed to the CIT/CTIO system ones through the equations in Persson et al. We estimate the uncertainties in the series of these system transformations should not exceed 0.05 mag. The photometry has a signal-to-noise ratio (S/N) of about 20 at $J$, $H$ and $K$ magnitudes of at 16.1, 16.1 and 15.0, respectively. Typical photometric errors are summarized in Table~\ref{typical}.
\begin{table}
\caption[]{Typical photometric errors}
\label{typical}
\begin{center}
\begin{tabular}{crrc}
\hline
\multicolumn{1}{c}{magnitude} & \multicolumn{1}{c}{$J$} & \multicolumn{1}{c}{$H$} & \multicolumn{1}{c}{$K$} \\
\hline
10.0 & 0.018 & 0.018 & 0.018 \\
11.0 & 0.018 & 0.018 & 0.018 \\
12.0 & 0.018 & 0.018 & 0.018 \\
13.0 & 0.018 & 0.018 & 0.020 \\
14.0 & 0.022 & 0.021 & 0.031 \\
15.0 & 0.030 & 0.029 & 0.049 \\
16.0 & 0.048 & 0.048 & ----- \\
\hline
\end{tabular}
\end{center}
\end{table}

In the present study, the data are not dereddened. Instead, we show the direction and magnitude of the reddening vector in each diagram, based on the relations in \citet{koornneef}.

\subsection{Detections of variable stars}
We used Image Subtraction Method (ISM) package ISIS.V2.1 (\citealt{alard1998}; \citealt{alard2000}) to detect variable stars. This method can find variable stars even in very crowded fields such as LMC bar centre, and its high efficiency in detecting variables is well known from the previous/ongoing surveys (MACHO, e.g., \citealt{alcock}; MOA, e.g., \citealt{bond}; OGLE, e.g., \citealt{alard1999}; etc.). In this study, variable stars with light variations of 0.03 mag or more would all have been detected. So far, nearly ten thousand variables have been detected in our surveying area, and their $JHK$ light curves have been accumulated.

\section{Results}
In the survey area, 717,239 sources were detected by DoPHOT. Because of the scan overlaps, there should be many multiple entries of a single source in this original sample. We eliminated such multiple entries based on (1) spatial proximity ($|\Delta r| \le 1.5^{''}$) and (2) difference in photometry ($|\Delta m_{J,H and K}| \le 0.1$ mag). It leaves 562,429 sources. Further, requiring detections at all wave bands with S/N$\ge$20 in $K$ leaves 226,793 sources. Among the 226,793 sources, DoPHOT classified 184,677 sources as `perfect' stars in all three wave bands (i.e., object type $=$ 1 in $J$, $H$ and $K$). Hereafter, we refer these 184,677 `perfect' stars as the main sample.

Among the main sample, 5,188 stars were found to be variables. The full discussions of their pulsation properties, such as pulsation period, amplitude, variability type and mean magnitude etc. will be made in the future paper. In this letter, we will concentrate only on the information that a star is variable or not.

\begin{figure*}
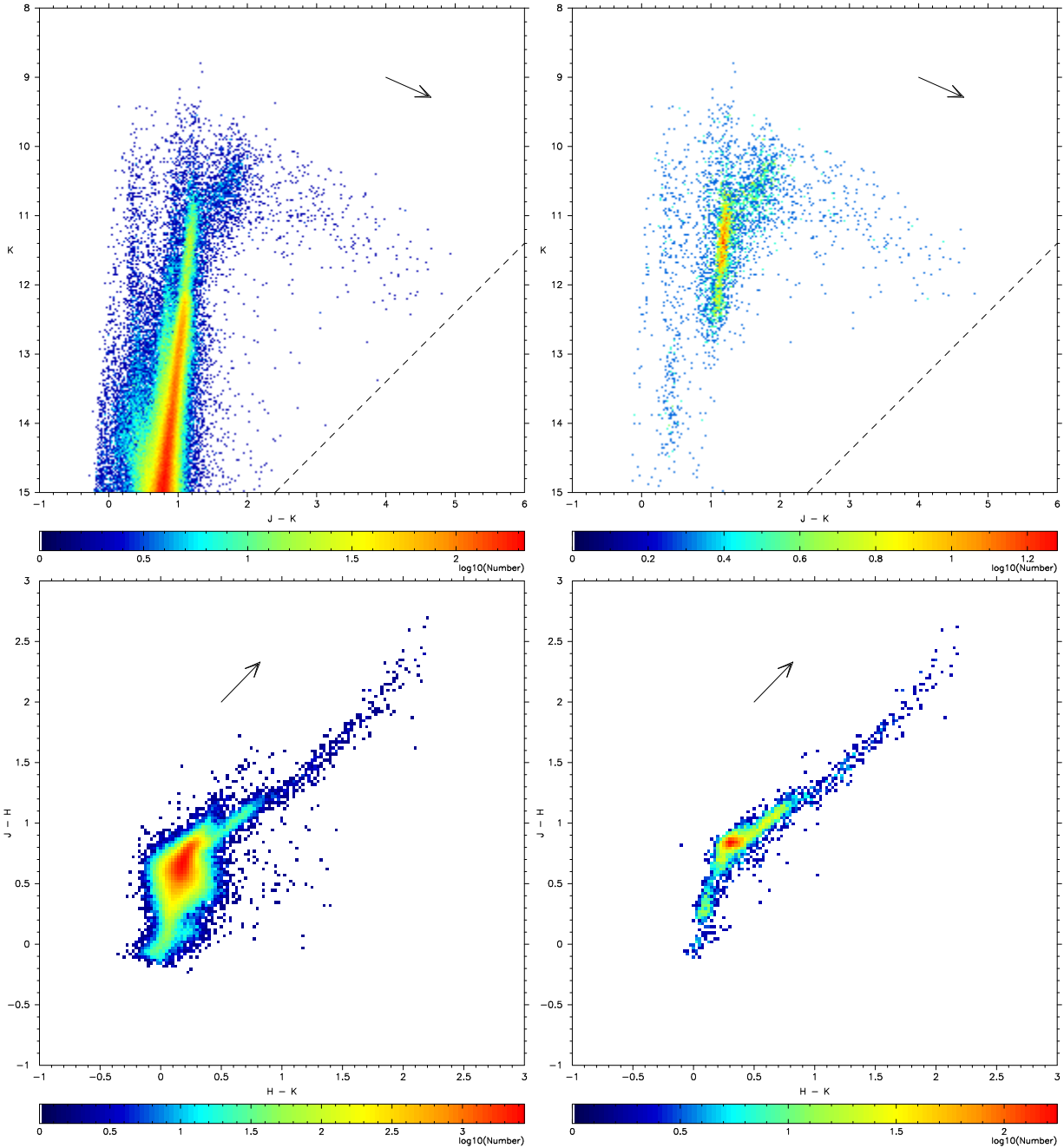

\centering
\includegraphics[angle=-90,scale=0.5]{yi2a.ps}
\includegraphics[angle=-90,scale=0.5]{yi2c.ps}
\includegraphics[angle=-90,scale=0.5]{yi2b.ps}
\includegraphics[angle=-90,scale=0.5]{yi2d.ps}
\caption{\textit{Left panel}: Colour-magnitude and colour-colour diagram of the 184,677 main sample stars. \textit{Right panel}: The same as the left panel, but of the 5,188 variables found in the main sample. Each plane is binned by 0.025 mag and the number density levels are logarithmic (see the wedge). The Reddening vector is drawn for $E_{B-V}=1.0$. A dashed diagonal line indicates the lower detection limit.}
\label{colmag}
\end{figure*}

\section{Discussion}
In Figure~\ref{colmag} left panel, we show the colour-magnitude diagram ($J-K, K$) and colour-colour diagram ($H-K, J-H$) of the main sample. In each diagram, numbers of stars in bins of area 0.025$\times$0.025 mag$^2$ are computed and the fiducial colour is applied according to the number density of stars in each position (see the annotated colour wedge). Because the plotted magnitudes are single-epoch ones, these are affected by the scatter due to the light variation of each source. As one can see, the lower right area and the upper part ($K \lesssim 9.5$) of the colour-magnitude diagram are sparsely populated. These are attributed to the $J$ limiting magnitude and $K$ band saturation of the SIRIUS, respectively.
 
\citet{nikolaev} analyzed the 2MASS LMC colour-magnitude diagram and identified the major stellar populations. Compared to their colour-magnitude diagram, ours has a narrow and sharp shape, suggesting a smaller Galactic foreground content in our sample. As this indicates, we can reasonably ignore the contribution of Galactic foreground stars here, because our survey region is centred on the LMC bar (see Figure.~\ref{map}), where foreground contribution is negligibly small compared to the number of the intrinsic LMC stars (\citealt{cioni2000a}).

From the result of our variable star survey, now we know which star is a variable. Since neither 2MASS nor DENIS provide variability information, it will be interesting if we add it to the colour-magnitude and the colour-colour diagram. Figure~\ref{colmag} right panel is the same diagram as the left one, but it only contains the 5,188 variables extracted from the main sample. By comparing left and right panels, we can tell which part of the diagram the variable stars reside in.

\subsection{Obscured variables}
Stars in the dust-enshrouded phase are particularly interesting because of their importance in many areas, such as in the supply source of interstellar matter and as a key to answer long standing problems, like the carbon star mystery (\citealt{iben1981}). Because of their redness, most, if not all, of these red variables have been possibly missed by the previous optical surveys, such as MACHO, MOA and OGLE. Previous observations of dust-enshrouded variables in the LMC (e.g., \citealt{zijlstra}; \citealt{reid}; \citealt{wood1992}; \citealt{vanloon}) were biased toward bright ones such as IRAS sources, and were carried out only for a few tens of stars. Therefore, it will be invaluable if their pulsation period, amplitude, and mean magnitudes would be known by our survey. We have to defer the full discussion of the properties of the obscured variables pending further variability observations, but it is of interest to show some preliminary results.

The Midcourse Space Experiment (MSX) satellite surveyed the whole LMC in the mid-infrared and many dust-enshrouded stars were detected (\citealt*{egan}; \citealt{wood2001}). These red stars are thought to be obscured by the dusty circumstellar envelope, which was developed by their mass-loss (\citealt{tanabe}). In the main sample, there are 254 very red ($J-K > 2.4$) stars. Most of them might be included in the MSX sources analysed by \citet{wood2001}. By comparing the left and right panel of Figure.~\ref{colmag}, one can easily notice the fact that most of such mass-losing stars are variables (194 stars out of 254 stars, 76.4\%). This observational result corroborates the idea that the mass-loss and variability are connected.

Some stars from \citet{hughes} have also very red colour and most of them are Carbon-rich. However, their sample only extends up to $J-K \lesssim 3.0$, and there is still a possibility that dusty Oxygen-rich variables like OH/IR stars (e.g., \citealt{wood1992}; \citealt{vanloon}) may have redder colours. Many obscured variables (up to $J-K \sim 5.0$) were found by our survey, but for most of them, the spectral type remains unknown. Infrared spectroscopic follow-up work would be interesting.

\subsection{RGB variables?}
In spite of the difficulty of distinguishing between the AGB and the RGB stars on the colour-magnitude diagram, variability is usually regarded as a sign of AGB stars. However, \citet{origlia} recently found a significant scale of mass-loss among the first giant branch stars near the RGB tip. Since a strong mass-loss is believed to be associated with pulsation, it is natural to search for the light variation in the RGB stars. 

\begin{figure}
\centering
\includegraphics[angle=0,scale=0.45]{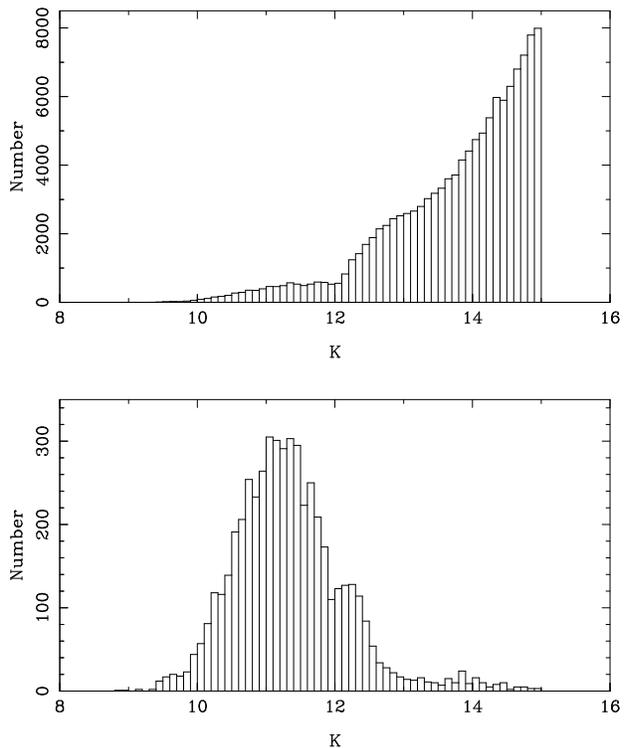}
\caption{Stellar $K$ magnitude distributions differentiated by 0.1 mag bins of main sample (upper panel) and of 5,769 variable stars extracted from it (lower panel). Note that the tip of the RGB is clearly seen.}
\label{lf}
\end{figure}

In the upper panel of Figure~\ref{lf}, we show the differential stellar $K$ magnitude distribution $N(m)$ ($dK =$ 0.1 mag) of the main sample. The discontinuity in the $N(m)$ is clearly seen around $K \approx 12.2$, which corresponds to the TRGB. Most of the stars lying above the TRGB are intermediate age AGB stars, while both RGB and AGB stars are present just under the TRGB with the ascendancy of the RGBs. \citet{renzini1992} suggested that in a composite population the $N(m)$ must exhibit a drop by more than a factor of 4 at the luminosity corresponding to the TRGB, and the ratio of AGB to RGB stars near the tip of the RGB is about 1/3 for the intermediate age stars. Our result confirms the predominancy of the intermediate population in the LMC.

We show the $K$ magnitude distribution for the 5,769 variables in the lower panel of Figure~\ref{lf}. As is clearly seen in it, there are two peaks in the luminosity of variables; one consists of genuine AGB variables and the other consists of stars just below the TRGB discontinuity. The position of the dip between the two peaks coincides with the above mentioned discontinuity for the main sample. 

\citet{alves} and \citet{wood1999} discussed the MACHO red variables that lie below the tip of the first giant branch. They suggested that all red giant variables are most likely AGB stars. According to \citet{origlia}, about 15\% of RGB stars near the TRGB have dust shells. Assuming all AGB stars vary and all of these 15\% of RGB stars near the TRGB also vary, and using Renzini's AGB/RGB ratio of 1/3, the ratio of variable RGB stars to AGB stars is at least 0.45. Although the second peak might be made of AGB variables with much fainter luminosities than the typical AGB variables, the coincidence of the dip luminosity for the variables with the discontinuity luminosity for the main sample can be more naturally explained by assuming that the second peak is due to a contribution from RGB variables since there is no reason to assume, according to models, that early AGB variables accumulate at the TRGB (\citealt{cioni2000b}).

\section{Summary}
In this letter, we presented the first results of our ongoing NIR survey of variable red giants in the LMC. Many obscured stars ($J-K > 2.4$) were detected and we found that most of them are variables, indicating that the mass-loss and variability are connected. The nature of the obscured variables are still unknown and further observational investigations are necessary and in progress. We also found that luminosity distribution of variable stars has two peaks. One is well above the TRGB and the other is just around it. Because there is no obvious reason to assume that Early AGB variables "pile-up" around the TRGB, we attributed the second peak to a contribution from RGB variables. From this point of view, we propose the evolutionary scenario of red giants as: (1) pulsate at the tip of the first giant branch; (2) pulsation fades away with the helium flash; (3) pulsation recommences as they ascend the asymptotic giant branch. However, the discrimination between RGB and AGB is still a challenging subject, and more direct evidence for their classification is needed.

\section*{Acknowledgments}
The authors would like to thank Michael Feast and John Menzies for a critical reading of the manuscript. We also thank Motonori Kamiya for kindly offering us a picture used in Figure~\ref{map}. This research is supported in part by the Grant-in-Aid for Scientific Research (C) No. 12640234 and Grant-in-Aids for Scientific on Priority Area (A) No. 12021202 and 13011202 from the Ministry of Education, Science, Sports and Culture of Japan. The IRSF/SIRIUS project was initiated and supported by Nagoya University, National Astronomical Observatory of Japan and University of Tokyo in collaboration with South African Astronomical Observatory under a financial support of Grant-in-Aid for Scientific Research on Priority Area (A) No. 10147207 of the Ministry of Education, Culture, Sports, Science, and Technology of Japan. Y. Nakajima, C. Nagashima and T. Nagayama are financially supported by Japan Society for the Promotion of Science. Finally, we thank the referee for kind and useful comments that helped to improve the paper.

%\bsp

\label{lastpage}

\end{document}